 \documentclass[authoryear,preprint,review,12pt]{elsarticle}

\usepackage{amssymb}

\begin{document}

\begin{frontmatter}



\title{Incomplete-exclusion Statistical Mechanics in Non-collisional Violent Relaxation of Celestial Objects}


\author{R. A. Treumann$^{a,c}$ \& W. Baumjohann$^b$}

\address[1]{Department of Geophysics, Munich University, Munich, Germany}
\address[2]{Space Research Institute IWF, Austrian Academy of Sciences, Graz, Austria}
\address[3]{International Space Science Institute, Bern}

\begin{abstract}
Violent relaxation has been proposed half a century ago to bear responsibility for  non-collisional 
dynamics and formation of gravitationally bound systems of extended celestial objects (agglomeration of stars, galaxies, clusters of galaxies) when reaching an approximate equilibrium state which can be described thermodynamically. The
Lynden-Bell equilibrium distribution of such systems, resulting from a spatial exclusion principle, had been shown to be an analog to the Fermi distribution of states in solid state physics. Real extended objects like galaxies do not completely exclude each other, however.
Permitting for partial exclusion leads to a modification of the equilibrium distribution. Here we show that this case can be
treated in analogy to a hypothetical incomplete population of Fermi states. An incomplete-exclusion equilibrium distribution
is obtained which enters the violent relaxation theory. 
\end{abstract}


\end{frontmatter}


\section{Introduction}
Fermi statistics got a new unexpected application when in 1967  \cite{lynden1967} suggested his ``principle of (volume) exclusion" in order to arrive at a statistical mechanical description of interacting stellar systems and formation of galaxies culminating in a so-called violent relaxation statistics of gravitationally bound non-collisional systems which presumably leads to a final ``thermodynamic equilibrium'' state. Subsequently, this violent relaxation theory found widespread application \cite[for a review see][]{gott1977} in galaxy formation and clusters of galaxies in particular in presence of cold dark matter \citep{kull2000}. It also found critical reviews \cite{shu1978} and improvements\citep{kull1996,kull1997}. The underlying idea was simply that two extended celestial bodies cannot at the same time occupy the same volume which Lynden-Bell called the ``exclusion principle''. Indeed, the two bodies necessarily exclude each other from occupying the same volume; otherwise they collide and may merge or disrupt -- all very complicated non-equilibrium processes outside any simple statistical mechanical description. 

Under Lynden-Bell's exclusion principle a volume-excluding statistical mechanics was constructed yielding the equilibrium distribution 
\begin{equation}\label{lb}
f_{LB}\propto \frac{\exp\{-\beta(\epsilon-\mu)\}}{1+\exp\{-\beta(\epsilon-\mu)\}}
\end{equation}
which formally resembles the Fermi distribution of zero-spin Fermions. Here $\epsilon$ is energy, $\beta,\mu$ are the Lagrange multipliers arising in the counting-of-states procedure and, as usual, being identified with an inverse temperature and a chemical potential, respectively.  Clearly, since the assumption of exclusion in the counting-of-states procedure is identical with the basic assumption of Pauli's exclusion principle underlying Fermi statistics, the above result is obvious -- even without derivation --, with the only difference that the proportionality factor is defined differently through the mass distribution of the interacting objects.  Using the above distribution in Poisson's equation $\nabla^2\psi=-4\pi G\int f_{LB}\mathrm{d}^3v$ for the gravitational potential $\psi$, with $v$ the velocity of gravitationally interacting `particles' (objects: stars, galaxies,...),  this yields Lynden-Bell's ``equation of violent relaxation'' for the potential of a gravitationally bound non-collisional system
\begin{equation}\label{eq-poisson}
\nabla^2\psi=-C\int\limits_0^\infty \frac{\zeta^2\mathrm{d}\zeta\exp(-\zeta^2/2)}{\exp[-\beta(\psi+\mu)]+\exp(-\zeta^2/2)}
\end{equation}
The integral on the right is essentially the mass density of the system under consideration, a self-consistent function of both: chemical potential $\mu$ and gravitational potential $\psi$. Integration is over the velocity space volume. The dimensional constant is $C=16\pi^2G\beta^{-\frac{3}{2}}\eta$, and $\eta$ is some unspecified initial phase space density, the normalization constant appearing in the above distribution. [The above distribution Eq. ({\ref{lb}) -- or its below derived equivalent Eq. (\ref{eq-fracferm}) -- can also be used in place of the equilibrium distribution in Vlasov theory in order to determine the linear eigenmodes of the violent relaxation process.] 

The assumption of complete volume exclusion is very strong, however, even for stars. Most stars have extended atmospheres, are magnetized and blow out winds. Hence, there is little reason to assume that in sufficiently dense though non-collisional compounds they strictly exclude each other from spatial overlap. As for an example, the extension of the Heliosphere of our Sun (the solar asterosphere), which is an isolated object, has radius of the order of 100-200 AU, with the heliosphere strongly interacting with the local interstellar matter. On this scale the Sun becomes a point-like object and the hypothetical interaction between it and another star would be mediated by the mixing of their asterospheres, long before any gravitational effects would come into play. In any system of non-isolated stars interaction would be much more violent. Even worse for galaxies in clusters of galaxies. Galaxies are extended objects which do by no means resemble solid bodies or point-like particles like electrons. They occupy large volumina and do necessarily spatially overlap when approaching each other, at least in their outskirts. Considering them as point-like objects in a statistical description is convenient but can only be justified as a rather weak approximation to reality. Modern approaches have turned to massive numerical simulations of large numbers of gravitating particles, en masse producing most relevant results of their temporal behavior. All such simulations, however, do not deviate from the assumption of non-extended objects; in this respect, considering their assumptions, they fall into the same category as the analytical statistical mechanical approach of Lynden-Bell \citep{lynden1967,lynden1979} and its refinements by  Shu \citep{shu1978} and others \citep{kull1996,kull1997}. Moreover, such simulations apply nicely to countable numbers of point-like particles. For large numbers like those encountered when simulating the entire universe they, however, become more uncertain, for each simulation particle plays the role of a macro-particle in the sense that it stands for an entire agglomeration of similar particles, the dynamics of which are assumed to behave not only similarly but identically. Relaxing the assumption of complete exclusion would be desirable therefore in all those cases though, for many reasons is difficult if not impossible to realize with the counting-of-states method for celestial objects, one of them the question of how to define an incomplete exclusion principle in a sufficiently general way comparable to Lynden-Bell's "complete" exclusion principle in the violent relaxation case. Fortunately, with this goal in mind we can exploit the analogy to Fermi statistics thereby circumventing the necessity of precisely including any overlap in state-counting statistics.

In the following we shall demonstrate how this can be done. We show below that the strict Pauli principle can be extended to the case of incomplete (or partial) occupation of states. In the statistical mechanics of elementary particles, to which Fermi statistics applies, the case of incomplete population of states is uncommon. Elementary particles like electrons either can or cannot be found in a particular state. The exempt are Anyons, collectively mixed fermionic and bosonic states obeying a statistics which exhibits properties of both Fermions and Bosons \citep{wilczek1982,haldane1983,haldane1991,halperin1984,wu1994}. They play a role in the fractional Quantum Hall effect \citep{wilczek1982,arovas1984}. Even though violent relaxation resembles Fermi statistics, Anyon statistics has not found an application in astrophysical systems yet for the reason of its complication. On the other hand, the "partially" -- or "incompletely" -- exclusive Fermi statistics we are going to derive is of a particularly simple form and should be more appropriate to application in the astrophysics of violent relaxation.

\section{Incomplete population of states}
There are three kinds of particles, Bosons, Fermions, and Anyons. As already noted, Anyons obey fractional statistics \citep{wilczek1982, haldane1983,arovas1984,wu1994} a mixture between bosonic and fermionic statistics. On the other hand, Bosons are allowed to occupy any energy states to arbitrary numbers; they do not resemble gravitationally interacting systems and thus need not be considered here. For Fermions, multi-particle states are inhibited by the Pauli principle which allows for only two occupations, empty states or (when neglecting the particle spin) one particle per state. Fermi's ingenious truncation of the infinite sum in the partition function 
\begin{equation}
Z_i=\sum_i\{\exp[(\mu-\epsilon_i)/k_BT]\}^{n_i} 
\end{equation}
by boldly assuming only binary occupations $n_i=[0,1]$ of states $\epsilon_i$, with $i\in\mathsf{N}$, immediately led him to the proposal  of the celebrated Fermi distribution $\langle n_i\rangle_F=\{1+\exp[-(\mu-\epsilon_i)/k_BT]\}^{-1}$. It had profound consequences for the statistical mechanics of solids at low temperatures. Fermi's assumption was justified by the Pauli principle and, for spin-$\frac{1}{2}$ particles, was ultimately given its quantum mechanical interpretation based on the complete asymmetry of fermionic wave functions. Discovery of the quantum Hall effect, in particular the fractional effect, had temporarily shaken Fermi statistics, until Laughlin's  \citep{laughlin1983} justifying proposal of his wave function which includes interaction with collective bosonic fields.  

One may ask what, on the other hand, happens, when sufficiently many states are available for Fermions (e.g. electrons) and the electrons would be allowed to fill such states only incompletely, e.g. in fractions? We may think of the above mentioned loosely interacting stars or galaxies, but for clarity we rather stay in the picture of having to deal with elementary particles, Fermions in our case. 

There are no experimentally known examples of incomplete filling of states for Fermions, since Fermions are assumed to either occupy a state or to be absent in this particular state. Such a population is binary. However, as for a theoretical example, one may for instance think of small numbers of gyrating electrons which bounce in a magnetic mirror geometry at frequency $\omega_b\ll\omega_{ce}$. In this case all Landau levels $\epsilon_L= \omega_{ce}\hbar (L +\frac{1}{2}), ~L\in \mathsf{N}$, split into a large number of bounce levels $\epsilon_b= \hbar\omega_b(b+\frac{1}{2}), ~ b\in\mathsf{N}, ~b/L<\omega_{ce}/\omega_b$. The total electron energy $\epsilon_{b,L}=\epsilon_L+\epsilon_b$ in Landau level $L$ is then shared by the two kinds of oscillatory states of the electron \citep{treu2013}. Under these circumstances, all the electron energy is in the Landau levels only at the magnetic mirror points, while in the minima of the magnetic field a substantial part of energy is transferred to the bounce levels. Theoretically, bounce levels might become partially (and thus incompletely) filled under these circumstances during the short gyration time, even though only fermionic states are involved, and partial population of states may not necessarily mean that the Pauli principle is violated, when not involving bosonic interactions. The
occupation may still be a fraction below one which the Pauli principle not explicitly excludes. Though this example may just serve for illustration, we simply assume that fermionic states can sometimes -- hypothetically -- become incompletely filled. In the following we rewrite the formalism for this case of partial (or incomplete) population, having in mind that it possibly is never realized in elementary particle statistical mechanics while becoming useful for some classical incompletely exclusive systems.

Rewriting Fermi statistics for the case of partially filled states is easily done, starting as usually \citep[cf., e.g.,][]{huang1987,landau1994} from the logarithm $\Omega_i[n_i]$  of the $\Gamma$-phase-space volume corresponding to the population numbers $[n_i]$ of the particles in an ideal gas:
\begin{equation}
\Omega_i = -k_BT\log \sum_{n_i} \left(\exp \frac{\mu -\epsilon_i}{k_BT}\right)^{n_i},
\end{equation}
i.e. proportional to the logarithm of the partition function $Z_i$, with $\mu$ chemical potential, $k_BT$ temperature, both in energy units, $\epsilon_i=p_i^2/2m$ particle energy, $\mathbf{p}$ particle momentum, $m$ mass, and the sum over Gibbs distributions in states $n_i$ the canonical partition function $Z_i$. 

Assume that the states can become partially occupied by Fermions alone. Occupation numbers $n_i>1$ are \emph{categorically  excluded} by the Pauli principle. Hence, partial population implies that, given the number interval $[0,\ell]$ with fixed natural number $\ell\in\textsf{N}$, and $j$ any integer such that $j\in[0,\ell]$, the thermodynamic potential $\Omega_i$ can be written
\begin{equation}
\Omega_i=-k_BT\log\sum_{j=0}^\ell\left(\exp\frac{\mu-\epsilon_i}{\ell k_BT}\right)^j, \qquad j\in[0,\ell]
\end{equation}
where $j=0,\ell$ just reproduces the two  permitted Fermi occupations. In the intermediate interval the populations follow the simple partial chain $\{j/\ell\}$, with fixed $\ell\geq j$. Summation of the sum becomes simple matter since it represents a truncated geometric progression with ratio $\exp[(\mu-\epsilon_i)/\ell k_BT]$ yielding
\begin{equation}
\Omega_i=-k_BT\log\left\{\frac{\exp[x_i(\ell+1)]-1}{\exp(x_i) - 1}\right\}, \qquad x_i\equiv \frac{\mu-\epsilon_i}{\ell k_BT}.
\end{equation}
One trivially shows that this becomes the thermodynamic Fermi potential $\Omega_{iF}=-k_BT\log\left\{1+\exp\left[(\mu-\epsilon_i)/k_BT\,\right]\right\}$ for $\ell=1$. From here, taking the derivative $-\partial\Omega_i/\partial\mu$, the average incompletely filled \emph{partial} distribution $\langle n_i\rangle_\ell$ in the $i$th quantum state follows as 
\begin{equation}
\langle n_i\rangle_\ell =\frac{1}{\ell}\left[\frac{(\ell+1)\mathrm{e}^{x_i(\ell+1)}}{\mathrm{e}^{x_i(\ell+1)}-1}-\frac{\mathrm{e}^{x_i}}{\mathrm{e}^{x_i}-1}\right],\quad \ell\geq 1,\label{eq-fracferm}
\end{equation}
It is again easily shown that the ordinary Fermi distribution $\langle n_i\rangle_F$ is reproduced for $\ell=1$. 

The  partial Fermi distribution Eq. (\ref{eq-fracferm}) is a bit more complicated than the (integer) Fermi distribution. 
Apparently, it looks more like a Boson distribution. However, this is an illusion which becomes clear when checking its low and high temperature forms which, as they should, agree with those for the Fermi distribution. For $T\to 0$ one obtains $\mu=\epsilon=\epsilon_F$, $\langle n_i\rangle\to 1$. For $T\gg\mu\sim\epsilon_F$ one recovers the Boltzmann distribution.

Thus the fermionic property of the distribution is maintained. Obviously it results from the subtraction of two bosonic distributions in Eq. (\ref{eq-fracferm}). The preserved anti-symmetric property of the partial many-particle fermionic system being caused by subtraction. 

From the partial distribution $\langle n_i\rangle_\ell$ as function of energy $\epsilon_i(\mathbf{p})$, which in its turn is a function of momentum $\mathbf{p}$, all thermodynamic quantities like the equation of state can be derived formally defining the appropriate moments. For the ideal Fermi gas one then has for the pressure $P$ and density $N$, respectively
\begin{equation}
\frac{P}{k_BT}=\frac{1}{\lambda_T^3}f_\ell(z), \qquad N=\frac{1}{\lambda_T^3}z\partial_zf_\ell(z)
\end{equation}
with $\log z = \mu/k_BT$, when introducing an appropriate new function
\begin{equation}\label{eq-fell}
f_\ell(z)=\frac{4}{\sqrt{\pi}}\int\limits_0^\infty y^2\mathrm{d}y\log\left(\frac{1-z^{1+\frac{1}{\ell}}\mathrm{e}^{-\left(1+\frac{1}{\ell}\right)y^2}}{1-z^{1/\ell}\mathrm{e}^{-y^2/\ell}}\right)\approx \ell^\frac{3}{2}\sum\limits_{k=1}^\infty \frac{(-1)^{k+1}z^{k/\ell}}{k^{5/2}}
\end{equation}
which replaces the usual function $f_{5/2}(z)$. Here, $\lambda_T=\sqrt{2\pi\hbar^2/mk_BT}$ is the thermal wavelength. As usual, ${\cal N}=z\partial_zf_\ell(z)$ is the particle number per volume spanned by the thermal wavelength.

\section{Continuous case} 
One may even formally extend the partial case to the extreme case of a continuity of partial states. Then $j/\ell$ becomes a continuous variable in the interval $j/\ell \in [0,1]$, and the sum in the expression for the thermodynamic potential $\Omega_i$ turns into an integral yielding that
\begin{equation}\label{contpot}
\Omega_i = -k_BT \log \left\{ \frac{k_BT}{\mu-\epsilon_i} \left[ \exp \left( \frac{\mu-\epsilon_i}{k_BT} \right) -1\right] \right\}.
\end{equation}
and for the average distribution
\begin{equation}\label{eq-fracferm-a}
\langle n_i\rangle = \frac{\exp\left[(\mu-\epsilon_i)/k_BT\right]}{\exp\left[(\mu-\epsilon_i)/k_BT\right]-1}-\frac{k_BT}{\mu-\epsilon_i}.
\end{equation}
Again the first term on the right looks like a Bose distribution. It is straightforward to show by expanding that, in the limit $k_BT\to 0$, it becomes
\begin{equation}\label{lyndcont}
\langle n_i\rangle_{T\to\,0} \simeq 1-\frac{k_BT}{\mu-\epsilon_i}\ \ {{\textstyle{-\!\!-\!\!\longrightarrow\atop{T=0}}}}\ \ 1, \qquad \epsilon_i<\mu,
\end{equation}
which is the completely degenerate Fermi distribution case. At large temperature the second term on the right cancels the first term, and the distribution algebraically approaches zero. This can already be seen from Eq. (\ref{contpot}) where for large $T$ one has $\epsilon_i>\mu$, in which case  the argument of the logarithm approaches unity. Consequently $\Omega_i\to 0$ for large temperature.


\section{Conclusions}
It is our aim to improve over the statistical mechanics underlying the theory of `violent relaxation'. Statistical mechanics of `violent relaxation' obeys some analogy to Fermi statistics. As an intermediate step we were thus asking for a modification of the Fermi statistics in order to include states which are incompletely filled. This might be a case which is not realized in the statistics of elementary particles, it however is necessary to account for incomplete exclusion of celestial bodies or systems. This intermediate step has been performed here. It results in the derivation of a `partial' Fermi distribution which has a particularly simple structure.   

Formally, there is no contradiction between the Pauli principle and incomplete filling of states. If a quantum state is filled just to a fraction of unity, i.e. a filling number smaller than one, this is not in disagreement with the Pauli principle which just restricts the number of particles in one state to not larger than one. This implies that `fractions of particles' -- if suitable and can be produced in some way -- are allowed to be found in a quantum state. Thus  from the purely statistical mechanical point of view there is no obvious contradiction between the Pauli principle for Fermions and any partial occupation of states. Statistical mechanics of Fermions, respecting the Pauli principle, does  not categorically exclude the existence of partially occupied states as long as the partial occupation number remains to be smaller than unity. 

The partial Fermi distribution Eq. (4) is a variant of the (integer occupation) Fermi distribution which it reproduces for $\ell=1$. The low and high temperature limits are the same as those of the Fermi distribution, as is the definition of Fermi energy. Fermions occupying partial states at $T=0$ remain to be degenerate, even for the case of continuous occupation of states below filling number one. As an interesting observation we note that for this type of distribution the fermionic character comes about by subtraction of two bosonic distributions. \footnote{One might believe that similar to the spatial exclusion principle in violent relaxation statistics the Pauli exclusion principle could result from some spatial exclusion of Fermions. However, this is not true. Fermions possess probability distributions in space and have no fixed positions which they could occupy at most in single numbers. The spatial  probability density clouds of a particular state would naively leave plenty of space for other Fermions to join. This, however, is strictly excluded by the Pauli principle.}

After this excursion into Fermi statistics we now return to the violent relaxation case. For incomplete exclusion of 'order $\ell'$ we may use the new distribution function Eq. (\ref{eq-fracferm}) which we write in the conventional form 
\begin{equation}\label{eq-dist1}
{\tilde f}_{\ell}\propto \frac{1}{\ell}\left\{\frac{\exp\left[-\beta(\epsilon-\mu)/\ell\right]}{1-\exp\left[-\beta(\epsilon-\mu)/\ell\right]}-\frac{(\ell+1)\exp\left[-(\ell+1)\beta(\epsilon-\mu)/\ell\right]}{1-\exp\left[-(\ell+1)\beta(\epsilon-\mu)/\ell\right]}\right\}
\end{equation}
The tilde indicates that we here are dealing with the Lynden-Bell distribution function which now is re-formulated for the incomplete exclusion case. The degree of incompleteness is contained in the parameter $\ell>1$, as explained before. The value of $\ell$ must be specified from observations, and when solving the violent relaxation Eq. (\ref{eq-poisson}), it is this Eq. (\ref{eq-dist1}) which must be used in the density integral. Since ${\tilde f}_\ell$ consists of two terms, the density integral also splits into two different terms:
\begin{eqnarray}
\nabla^2\psi=-&{\tilde C}&\int\limits_0^\infty\zeta^2\mathrm{d}\zeta\left\{\frac{\exp\left(-\zeta^2/2\ell\right)}{\exp\left[-\beta(\psi+\mu)/\ell\right]-\exp\left(-\zeta^2/2\ell\right)}-\right.\\ 
&-&\left.\frac{(\ell+1)\exp\left[-(\ell+1)\zeta^2/2\ell\right]}{\exp\left[-(\ell+1)\beta(\psi+\mu)/\ell\right]-\exp\left[-(\ell+1)\zeta^2/\ell\right]}\right\}
\end{eqnarray}
and ${\tilde C}\equiv C/\ell$. This equation must be solved numerically for any particular object applying some Poisson solver. By fitting the data to the solution one might be able to determine $\ell$, the ``degree of exclusion", in our language.

When a continuous spectrum of states is assumed, a whole continuous spectrum of overlaps should become possible. Then the continuous distribution Eq. (\ref{lyndcont}) is responsible. One realizes that its two parts are just the sum of the Bose form of Lynden-Bell's exclusion distribution which is corrected by the second term to account for the fermionic properties. At large $\beta$ (zero temperature) the problem becomes the completely degenerate case.   For small $\beta$ (large temperature) the two terms compensate each other, as can be checked by expanding. Thus the distribution vanishes identically. There is no continuous distribution of incompletely filled states at large temperature. Accordingly, using the continuous distribution in Poisson's equation of violent relaxation yields
\begin{equation}
\nabla^2\psi= -C\int\limits_0^\alpha\zeta^2\mathrm{d}\zeta\left\{\frac{1}{\beta(\psi+\mu)+\zeta^2/2}-\frac{\exp(-\zeta^2/2)}{\exp\left[-\beta(\psi+\mu)\right]-\exp(-\zeta^2/2)}\right\}
\end{equation}
In spite of the vanishing of the thermal mass distribution, the integral of the first term with infinite upper limit $\alpha\to\infty$ does not exist and is, moreover, not compensated by the second term. The integral thus would make sense only up to a finite upper energy limit $\alpha\ll\infty$ thereby indicating that for arbitrary exclusion modes at high energies and temperatures the assumption of spatial exclusion will become violated and other effects like collisions and merging must be taken into account and cannot be neglected a priori. This implies that the systems will become collisional, when relaxation is different. 

\section{Summary}
We have fulfilled our promise to improve over the strict exclusion principle underlying violent relaxation. This was possible by constructing a \emph{hypothetical} partial-filling theory of fermionic states in Fermi statistical mechanics which, as we pointed out, does not violate the Pauli principle because the latter does not categorically inhibit incomplete filling of states. It actually preserves the Pauli principle -- one may note that a generalization of the Pauli principle has been proposed for Anyons \citep{haldane1991}. We do not claim that this kins of statistics which extends Fermi statistics to incomplete state occupation will find an application in elementary particle physics of Fermions. This was not intended; we aimed for a partial filling of states in Fermi statistics because of its analogy to exclusion statistics in violent relaxation theory. This analogy has been exploited in view of an input into violent relaxation theory with the philosophy in mind that extended object do by no means exclude each other completely from the volume each of them occupies. At the contrary the exclusion is incomplete, and there will be a substantial mixing at least in the outskirts of the objects. This applies possibly less to stars but is expected for the interaction of galaxies in clusters of galaxies and might affect their gravitational structure. Incomplete filling of states in Fermi statistics implies the existence of discrete states which are characterized by a fraction number $\ell$ which has a clear physical meaning. In the analogy of violent relaxation this meaning is lost; here $\ell$ characterizes the `state of exclusion' which is a quantitative classification the qualitative meaning of which must be elucidated from application of violent relaxation theory to real objects. This lies outside of the present paper. It requires solution of the equation for the gravitational potential and fitting to observational data. 

We also derived the distribution for a hypothetical state of a continuum of states which are partially populated, corresponding to a continuum of exclusion states in violent relaxation. Here it turns out that the equation for the gravitational potential in violent relaxation contains a diverging term, which suggests that cases like this may not be possible to realize in violent relaxation theory. It is not entirely clear what this means in this context. One may, however, suggest, that a continuum of exclusions, assuming only gravitational interactions, possibly overbuys gravitational theory by neglecting many other effects. It seems that the violent gravitation equation cannot deal with this case which probably requires inclusion of other possibly more violent processes like collisions.   

\subsection*{Acknowledgement}
{Hospitality of the International Space Science Institute, Bern, is acknowledged.}



\end{document}